\documentclass[%
superscriptaddress,
 amsmath,assume,
 aps,
prl,
twocolumn,
]{revtex4-2}

\usepackage{multirow}
\usepackage{graphicx}
\usepackage{dcolumn}
\usepackage{bm}
\usepackage[caption=false]{subfig}
\usepackage{stackengine}
\usepackage{tikz}
\usepackage{array}
\usepackage{multirow}
\usepackage{nccmath}

\usepackage{color}

\begin{document}


\title{Quantifying $\alpha$ clustering in the ground states of $^{16}$O and $^{20}$Ne}

\author{E. Harris}
 \affiliation{Department of Physics \& Astronomy, Texas A\&M University, College Station, Texas 77843, USA}
 \affiliation{Cyclotron Institute, Texas A\&M University, College Station, Texas 77843, USA} 
\author{M. Barbui}%
 \affiliation{Cyclotron Institute, Texas A\&M University, College Station, Texas 77843, USA} 
\author{J. Bishop}
 \altaffiliation{Current Address: School of Physics and Astronomy, University of Birmingham, Edgbaston, Birmingham, B15 2TT, United Kingdom}
 \affiliation{Cyclotron Institute, Texas A\&M University, College Station, Texas 77843, USA} 
\author{G. Chubarian}
 \affiliation{Cyclotron Institute, Texas A\&M University, College Station, Texas 77843, USA}
\author{Sebastian König}
\affiliation{Department of Physics, North Carolina State University,
Raleigh, NC 27695, USA}
\author{E. Koshchiy} 
\thanks{Deceased}
 \affiliation{Cyclotron Institute, Texas A\&M University, College Station, Texas 77843, USA}
\author{K.D. Launey}
\affiliation{Department of Physics \& Astronomy, Louisiana State University, Baton Rouge, Louisiana 70803, USA}
\author{Dean Lee} 
\affiliation{Department of Physics \& Astronomy, Michigan State University, East Lansing, Michigan 48824, USA}
\affiliation{Facility for Rare Isotope Beams (FRIB), Michigan State University, East Lansing, Michigan 48824, USA}
\author{Zifeng Luo}
\affiliation{Department of Physics \& Astronomy, Texas A\&M University, College Station, Texas 77843, USA}
 \affiliation{Cyclotron Institute, Texas A\&M University, College Station, Texas 77843, USA}
\author{Yuan-Zhuo Ma}
 \affiliation{Facility for Rare Isotope Beams (FRIB), Michigan State University, East Lansing, Michigan 48824, USA}
 \author{Ulf-G. Mei{\ss}ner}
 \affiliation{Helmholtz Institut f\"ur Strahlen- und Kernphysik and Bethe Center for Theoretical Physics,\\ Universit\"at Bonn, D-53115 Bonn, Germany}
\affiliation{ Institute for Advanced Simulation (IAS-4) Forschungszentrum J\"ulich, Germany}
\author{C.E. Parker}
 \affiliation{Cyclotron Institute, Texas A\&M University, College Station, Texas 77843, USA}
 \author{Zhengxue Ren}
\affiliation{Institut~f\"{u}r~Kernphysik,~Institute~for~Advanced~Simulation,~
J\"{u}lich~Center~for~Hadron~Physics,\protect\linebreak Forschungszentrum~J\"{u}lich,
~D-52425~J\"{u}lich,~Germany}
\affiliation{Center for Advanced Simulation and Analytics (CASA),~Forschungszentrum~J\"{u}lich,~D-52425~J\"{u}lich,~Germany}
\author{M. Roosa}
 \affiliation{Department of Physics \& Astronomy, Texas A\&M University, College Station, Texas 77843, USA}
 \affiliation{Cyclotron Institute, Texas A\&M University, College Station, Texas 77843, USA}
\author{A. Saastamoinen}
 \affiliation{Cyclotron Institute, Texas A\&M University, College Station, Texas 77843, USA}
 \author{G. H. Sargsyan}
 \affiliation{Facility for Rare Isotope Beams (FRIB), Michigan State University, East Lansing, Michigan 48824, USA}
\author{D.P. Scriven}
 \affiliation{Department of Physics \& Astronomy, Texas A\&M University, College Station, Texas 77843, USA}
 \affiliation{Cyclotron Institute, Texas A\&M University, College Station, Texas 77843, USA}
\author{Shihang Shen}
\affiliation{Institut~f\"{u}r~Kernphysik,~Institute~for~Advanced~Simulation,~
J\"{u}lich~Center~for~Hadron~Physics,\protect\linebreak Forschungszentrum~J\"{u}lich,
~D-52425~J\"{u}lich,~Germany}
\affiliation{Center for Advanced Simulation and Analytics (CASA),~Forschungszentrum~J\"{u}lich,~D-52425~J\"{u}lich,~Germany}
\author{A. Volya}
\affiliation{Department of Physics, Florida State University, Tallahassee, Florida 32306, USA}
\author{Hang Yu}
\affiliation{Department of Physics, North Carolina State University,
Raleigh, NC 27695, USA}
\affiliation{Center for Computational Sciences, University of Tsukuba, Tsukuba, Ibaraki 305-3577, Japan}
\author{G.V. Rogachev}
  \email{rogachev@tamu.edu}
 \affiliation{Department of Physics \& Astronomy, Texas A\&M University, College Station, Texas 77843, USA}
 \affiliation{Cyclotron Institute, Texas A\&M University, College Station, Texas 77843, USA} 
 \affiliation{Nuclear Solutions Institute, Texas A\&M University, College Station, Texas 77843, USA}
 
\date{\today}

\begin{abstract}
Understanding the role of multi-nucleon correlations in the structure of light nuclei is at the forefront of modern nuclear science. In this letter, we present a quantitative benchmark study of $\alpha$-cluster correlations in the ground states of $^{16}$O and $^{20}$Ne. Experimental data provide direct evidence that the wave functions of the ground states of $^{16}$O and $^{20}$Ne are dominated by $\alpha$-cluster correlations, in agreement with the predictions of sophisticated nuclear structure models. We also provide a new model-independent constraint for the $\alpha$ asymptotic normalization coefficient of the $^{16}$O ground state and discuss the implications of these findings on the $^{12}$C($\alpha$,$\gamma$)$^{16}$O reaction, which is of critical importance for nuclear astrophysics.
\end{abstract}

\maketitle

{\it Introduction}---$\alpha$ clustering phenomena in nuclei have profound impact on nucleosynthesis in stellar environment. Abundance of helium in the universe makes nuclear reactions with $\alpha$-particles central to nucleosynthesis, especially in the later stages of stellar evolution, after hydrogen fuel is exhausted. These reactions critically depend on the $\alpha$-cluster aspects of nuclear structure. The classical example of such a reaction is $^{12}$C($\alpha$,$\gamma$)$^{16}$O, which determines the ratio of carbon to oxygen in the universe. The cross section of this reaction at the Gamow window [near 300 keV in center of mass (c.m.)] is determined by the $\alpha$-cluster structure of the ground and excited states in $^{16}$O \cite{deBoer2017}.

Numerous experiments and theoretical studies indicate that multi-nucleon correlations or clusters play an important role in nuclear structure (see recent reviews \cite{Freer2018}, \cite{Stuchbery2022} and \cite{Lombardo2023}). A beautiful pattern of excited states exhibiting strong multi-nucleon correlations is observed across a wide range of nuclear reactions and resonance scattering. For example, above the $\alpha$-decay threshold, the $\alpha$-cluster states manifest themselves as strong resonance peaks in the excitation function of resonance elastic scattering of $\alpha$-particles on certain target nuclei \cite{Goldberg1997}. The most direct evidence for the significance of cluster configurations in these studies is provided by the large partial $\alpha$-widths of the observed resonances. These widths are typically compared to extreme values, which can be evaluated for purely core+$\alpha$ systems using a simple potential model. In particular, some excited states in $^{16}$O and $^{20}$Ne are known to have strong cluster configurations in this sense. For instance, the partial $\alpha$-widths of the  1$^-$ states at 9.59~MeV in $^{16}$O \cite{Tilley1993} and at 5.79~MeV in $^{20}$Ne \cite{Tilley1998} approach the extreme values predicted by core+$\alpha$ potential models, indicating well-developed $\alpha$-cluster structures.

The quantification of the degree of clustering for states below the relevant decay threshold is more ambiguous. For $\alpha$-cluster configurations, this is typically achieved through $\alpha$-transfer reactions, such as ($^6$Li,d) or ($^7$Li,t), where $\alpha$-spectroscopic factors (SFs) are extracted from the measured reaction cross sections and reaction model analysis (e.g., Distorted Wave Born Approximation, DWBA). However, these results are model-dependent, influenced by choices such as the optical model potentials (OMPs), the shape of the form-factor potentials, and other model parameters. One way to reduce model dependence is by extracting Asymptotic Normalization Coefficients (ANCs) \cite{Mukhamedzhanov2022}. ANCs are observables that represent the amplitude of the bound-state wave function in the asymptotic region, making them model-independent and directly measurable in the peripheral nuclear reactions. Performing the $\alpha$-transfer nuclear reaction at sub-Coulomb energies leads to near elimination of the model dependences and the absolute values of the $\alpha$ ANCs can be established this way. This approach, pioneered by C. Brune \cite{Brune1999}, has been successfully applied to a number of nuclei \cite{Johnson2006,Johnson2009}. The main advantage of sub-Coulomb $\alpha$-transfer reactions is the significantly reduced role of the OMPs due to the highly peripheral nature of the reaction. It is crucial that both the entrance and exit channels lie below the Coulomb barrier in this approach. As a result, only a subset of states within a narrow range near the $\alpha$-decay threshold can be probed this way. The goal of this work is to develop a model-independent experimental technique to quantify $\alpha$-clustering in ground states for benchmarking the predictions of the state-of-the-art theoretical calculations and for nuclear astrophysics applications. By applying it to $^{16}$O and $^{20}$Ne, we provide new model-independent constraint for the $^{16}$O ground state $\alpha$-ANC and explore its impact on the $^{12}$C($\alpha$,$\gamma$)$^{16}$O reaction rate.

The advancement of nuclear theory in quantifying clustering in atomic nuclei over the last two decades has been remarkable. We made a transition from the early models that assumed presence of clusters {\it a priori} to highly sophisticated microscopic approaches which do not make any assumptions about the multi-nucleon correlations and often use realistic nucleon-nucleon and multi-nucleon interactions. An early example of a model with \emph{a priori} clustering assumptions was suggested by Robson \cite{Robson1979} and then refined by Iachello \cite{Iachello1982} in the framework of the Algebraic Cluster Model (ACM). This model applies symmetry considerations and assumes that four-nucleon correlations ($\alpha$-clusters) play a dominant role in some nuclei, of which $^{12}$C and $^{16}$O are prime examples. Remarkably, the most sophisticated microscopic theoretical approaches available today, such as the  Fermionic Molecular Dynamics approach (FMD) \cite{Chernykh2007} and Nuclear Lattice Effective Field Theory \cite{Epelbaum2012}, confirm, at least partially, the validity of the assumptions of the early models; similarly, $\alpha$-clustering features have been identified in the symmetry-adapted no-core shell-model framework for the ground states of $^8$Be \cite{LauneyDD16} and $^{20}$Ne \cite{DreyfussLESBDD20,SARGSYAN2025} (see also \cite{Launey2024} for $^{16}$O). It is becoming increasingly clear that multi-nucleon correlations play a fundamental role not only in the excited but also ground states of light nuclei. 

\begin{figure}
    \centering
    \includegraphics[width=\linewidth]{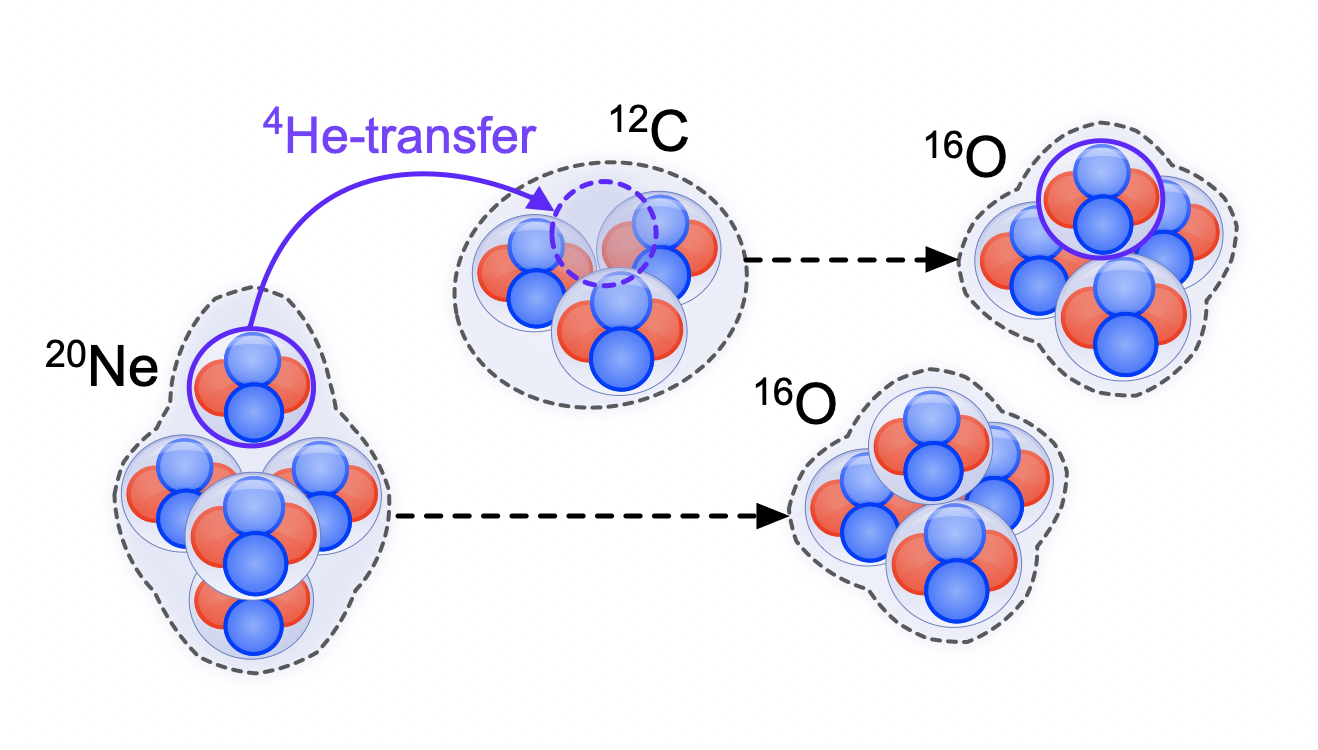}
    \caption{Illustration of the $^{12}$C($^{20}$Ne,$^{16}$O)$^{16}$O $\alpha$-transfer reaction.}
    \label{fig:transfer_diagram}
\end{figure}

Probing such correlations experimentally in a quantitative way is challenging. Focusing on the $^{16}$O and $^{20}$Ne ground states and aiming at reducing the model-dependence of the results as much as possible, we used the following approach. First, we identify the most model-independent reaction and perform measurements under the conditions which reduce the systematic uncertainties as much as possible. We have selected the $^{12}$C($^{20}$Ne,$^{16}$O)$^{16}$O reaction performed at energies around 1 MeV/nucleon as a probe for the g.s. of $^{16}$O and $^{20}$Ne (Fig. \ref{fig:transfer_diagram}). At this energy the $\alpha$-transfer is highly peripheral and therefore is nearly independent of the OMPs. We extract the product of the squares of the $\alpha$-ANCs of the $^{16}$O and $^{20}$Ne ground states, which is directly proportional to the reaction cross section and is nearly independent of the assumptions of the reaction theory with respect to the shape of the form-factor potentials or the number of nodes in the cluster wave-functions. 

Simultaneously, we perform three independent theoretical studies to calculate the square product of the $\alpha$-ANCs of the $^{16}$O and $^{20}$Ne ground states. We benchmark to three sophisticated microscopic approaches: Nuclear Lattice Effective Field Theory (NLEFT) \cite{Epelbaum2012}, Symmetry-Adapted No-Core Shell Model (SA-NCSM) \cite{LauneyMD_ARNPS21}, and Cluster Nucleon Configuration Interaction (CN-CI) \cite{Kravvaris2019}. The predictions of these studies for the $^{20}$Ne $\alpha$-ANC are then used to extract the $^{16}$O(g.s.) $\alpha$-ANC from our experimental data. Further details about these models are provided in the supplemental material and references therein.

An important feature of this study was that the experimental work and the analysis of the experimental data was performed without any knowledge of the results of the theoretical calculations and, vice versa, the theoretical calculations were performed without any knowledge of the experimental results and/or the results of the calculations from the competing approaches. The collaboration asked an independent observer, the Director of the Cyclotron Institute at Texas A\&M University, Dr. S. Yennello, to collect the results from all four groups (one experimental and three theoretical) and to unblind the results to the entire collaboration simultaneously. This approach prevented unconscious confirmation bias in this benchmark study.  
\\

{\it Experiment}---The $^{12}$C($^{20}$Ne,$^{16}$O)$^{16}$O measurement was conducted at the Texas A\&M University Cyclotron Institute using a beam of $^{20}$Ne at 1.0~MeV/u (20~MeV total kinetic energy) produced by the K150 cyclotron. The beam was impinged on a natural carbon target with a measured thickness of 22 $\mu$g/cm$^2$. This configuration corresponds to a c.m.\ energy of 7.5~MeV for the $^{20}$Ne+$^{12}$C system, which is below the Coulomb barrier of $\sim$11 MeV. At this sub-Coulomb energy, the $\alpha$-transfer cross section is maximum at large backward angles in the c.m.\ frame, and by utilizing inverse kinematics, we create favorable conditions for reaction products to be directed at small forward angles in the laboratory with reasonable energies for detection. 

The Multipole-Dipole-Multipole (MDM) magnetic spectrometer, placed at an angle of 5$^{\circ}$ in the laboratory frame, was employed to filter events of interest based on their magnetic rigidity. Detection of the $^{16}$O $\alpha$-transfer events, after being filtered by the MDM, is done with the Texas Parallel-Plate Avalanche Counter (TexPPAC), the new MDM focal plane detector developed for measuring low energy ($\sim$1 MeV/u) heavy ions. It features two large area, position sensitive PPAC detectors spaced 42 cm apart within a single gas chamber filled with 4 Torr of Pentane gas. Systematic uncertainties of the result are eliminated by measuring ratios of the reaction cross section to $^{12}$C+$^{20}$Ne elastic scattering and comparing these ratios directly to the reaction model predictions. 

The Distorted Wave Born Approximation (DWBA) calculations using FRESCO \cite{Thompson1988} were performed to evaluate the reaction cross sections. We used phenomenological optical potentials from \cite{Vandenbosch1974} for the $^{12}$C+$^{20}$Ne channel and the $^{12}$C+$^{16}$O core-core potential, and  two sets of potentials from \cite{Wu1984} for the $^{16}$O+$^{16}$O channel. Both $\alpha$+$^{16}$O and $\alpha$+$^{12}$C real binding potentials adopt a Woods-Saxon form factor. As expected, the results are not sensitive to the shape of these potentials, as well as on other parameters of the calculations, highlighting the advantages of using sub-Coulomb $\alpha$-transfer reaction. The overall theoretical uncertainty of our analysis is 20\%. It is  primarily due to dependence of the DWBA calculations on the parameters of the exit channel optical model potential ($^{16}$O+$^{16}$O, see Fig. \ref{fig:cs_ratio} in supplemental materials).

{\it Results and Discussion}---The experimental result and a summary of the theoretical $\alpha$ ANCs from the NLEFT, SA-NCSM, and CN-CI models are shown in Table \ref{theoretical_ANCs}. Both the individual ANCs and the product of the squared ANCs compared to the experimental value are shown for each model, with further information on each calculation detailed in the supplemental material.


\begin{table}[ht]
\setlength{\extrarowheight}{4pt}
\centering
\begin{tabular}{lccc}
\hline
\\
 Experiment & \multicolumn{3}{c}{$(\text{C}^{^{16}\text{O}}_{\alpha,^{12}\text{C}})^2(\text{C}^{^{20}\text{Ne}}_{\alpha,^{16}\text{O}})^2 = 9.4(1.4)(2)\times10^{12}~\text{fm}^{-2}$} \\
 \\
 \hline
 &  $\text{C}^{^{16}\text{O}}_{\alpha,^{12}\text{C}}$ & $\text{C}^{^{20}\text{Ne}}_{\alpha,^{16}\text{O}}$ & $[(\text{C}^{^{16}\text{O}}_{\alpha,^{12}\text{C}})^2(\text{C}^{^{20}\text{Ne}}_{\alpha,^{16}\text{O}})^2]^{1/4}$ \\
 & (fm$^{-1/2}$) &  (fm$^{-1/2}$) & ($10^3$ fm$^{-1/2}$) \\
\hline
Experiment    &  &  & 1.75(7)(9)\\
NLEFT& 380(80)& 3800(950)& 1.20(25)\\
SA-NCSM  & 210(30)& 3760(470)& 0.89(14)\\
CN-CI & 520(30) & 4500(500)& 1.53(13)\\ 
Extreme CM & 640& 5700& 1.9 \\
\hline
\end{tabular}
\caption{Theoretical $\alpha$-particle ANCs for the $^{16}$O and $^{20}$Ne ground states and their geometric mean, compared to the respective experimental value. The uncertainty of the experimental value is given in order of statistical and theoretical systematic error.}
\label{theoretical_ANCs}
\end{table}

Given the minimal dependence of our experimental results on the model parameters, the large value of the square of the product of the ANCs, $(\text{C}^{^{20}\text{Ne}}_{\alpha,^{16}\text{O}})^2(\text{C}^{^{16}\text{O}}_{\alpha,^{12}\text{C}})^2$, provides direct evidence of the key role of $\alpha$-cluster correlations in the ground states of $^{16}$O and $^{20}$Ne. Before discussing a benchmark to the microscopic calculations, it is instructive to compare the experimental results to the predictions of a simple $\alpha$+core potential model. Within this model, the ground states of $^{16}$O and $^{20}$Ne can be represented by a wave function describing the relative motion of the $\alpha $-cluster and the core. The $\alpha$-core interaction can be approximated by the Woods-Saxon potential, with shape parameters (radius and diffuseness) adjusted to reproduce the partial $\alpha$-width of the 1$^-$ cluster states in $^{16}$O and $^{20}$Ne, as mentioned in the introduction. The depth of the potential is fit to reproduce the $\alpha$ binding energies while maintaining the correct number of nodes in the cluster wave function to obey the Pauli principle (two for $^{16}$O and four for $^{20}$Ne). This simple model gives $\sim$ 13$\times 10^{12}$~fm$^{-2}$ for the square product of the ANCs. Note that the experimental result is only 30\% below this extreme model.

To quantitatively compare to the experimental value, we use the geometrical mean $(\text{C}^2\text{C}^2)^{1/4}$, which retains the ANC scale. Overall, there is an agreement among the different theoretical approaches within 1-2 sigmas. Further, achieving predictions within a factor of 1.2-2 from the experimental result for highly non-trivial $\alpha$-ANCs calculations showcases the capabilities of modern microscopic approaches.

The variation in the theoretical predictions for the $^{16}$O ANC may be due to the relatively compact structure of $^{16}$O which makes it particularly sensitive to details of its nuclear structure. In contrast, the results indicate a general agreement among $^{20}$Ne ANCs within uncertainties. We note that the $\alpha$ separation energy for $^{16}$O is $7.16$~MeV, while the $\alpha$ separation energy for $^{20}$Ne is only $4.73$~MeV.  The more loosely-bound structure of $^{20}$Ne, which in turn exhibits a more pronounced surface $\alpha$ clustering, makes $^{20}$Ne ANC calculations more robust against systematic errors, provided that the physical separation energy is well reproduced. Further refinement of these calculations may be necessary to reconcile the observed variations in $^{16}$O ANCs (see Refs.~\cite{Giacalone:2024luz,LauneyMD_ARNPS21} regarding the cluster structures of $^{16}$O and $^{20}$Ne). The consistency of $^{20}$Ne ANC results allow us to use the $^{20}$Ne ANCs and the experimental result to make a prediction of the $^{16}$O ground state $\alpha$-ANC, which is a crucial parameter used in constraining the $^{12}$C($\alpha,\gamma$)$^{16}$O reaction \cite{deBoer2017}.

\begin{figure}
\includegraphics[width=\linewidth]{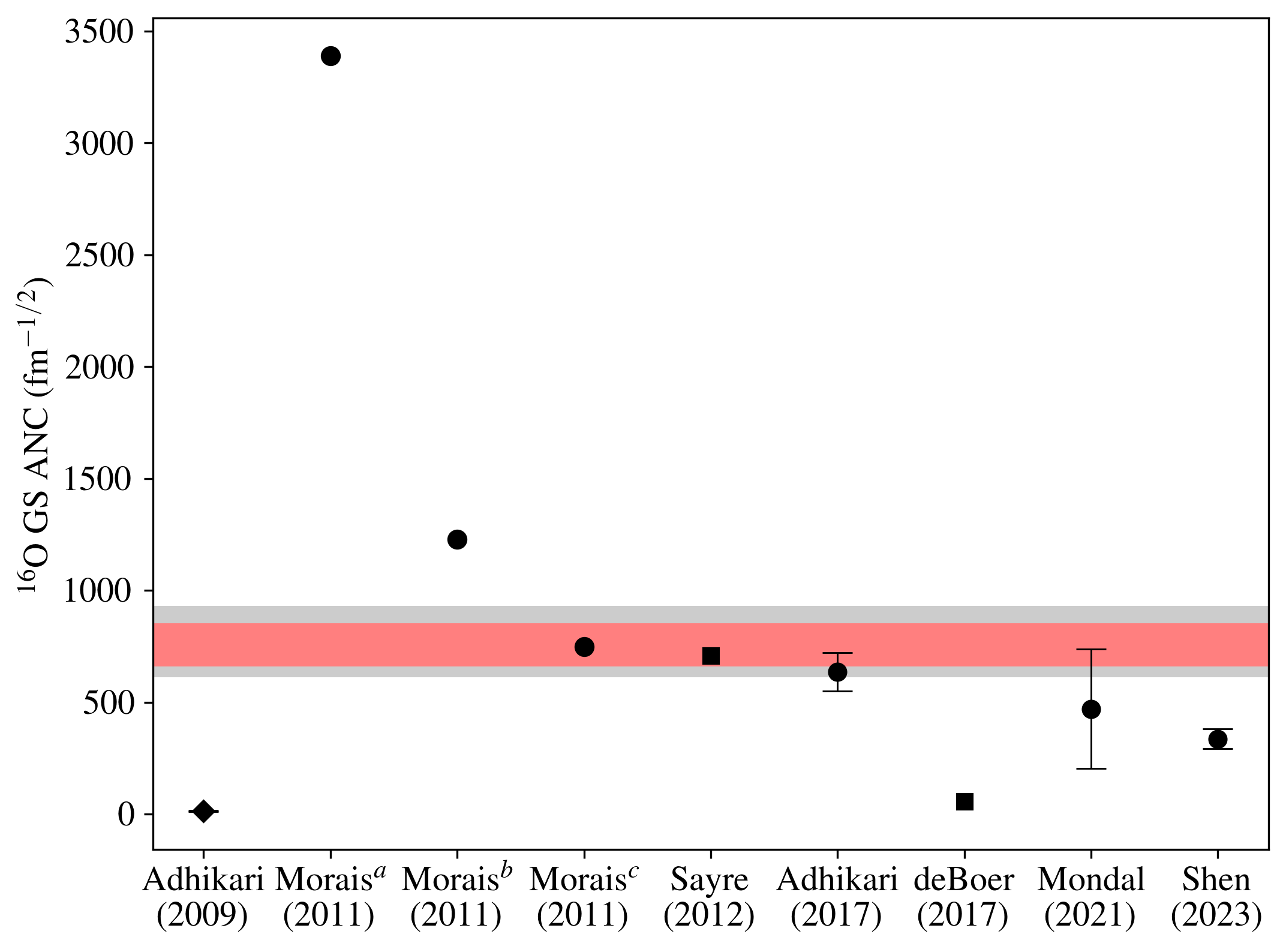}
    \caption{Comparison of $^{16}$O ground state $\alpha$-ANC values available in literature with uncertainty bands representing the range of possible ANC values determined by this work.  Results taken from Adhikar \cite{Adhikari2009}, Morais \cite{Morais2011}, Sayre \cite{Sayre2012}, Adhikari \cite{Adhikari2017}, deBoer \cite{deBoer2017}, Mondal \cite{Mondal2021},  and Shen \cite{Shen2023}. The grey band represents conservative uncertainties from the microscopic calculations, while the red band corresponds to experimental uncertainties only. The black circles represent measurements conducted using transfer reactions, the black diamond represents a measurement conducted using $^{16}$O breakup, and the black squares are from R-matrix evaluations.}
    \label{fig:16O-ANCs}
\end{figure}

Several attempts have been made to measure the $^{16}$O ground state ANC due to its significance in nuclear astrophysics, but there exists a large discrepancy between results obtained by various methods, including R-matrix analysis \cite{Sayre2012}, transfer reactions \cite{Morais2011, Adhikari2017, Mondal2021, Shen2023}, and breakup reactions \cite{Adhikari2009}. In this work, we provide limits on the $^{16}$O ANC value by applying the three theoretical $^{20}$Ne values to our experimental result. The overall effect of these constraints on the $^{16}$O ground state ANC is demonstrated in Fig. \ref{fig:16O-ANCs}. The conservative upper and lower limits, represented by the grey band, encompass the theoretical uncertainties from the calculations, while the red bad includes only experimental uncertainties as quoted in the results of Table \ref{theoretical_ANCs}. Our constraints for $^{16}$O(g.s.) $\alpha$-ANC, obtained from the experimental value of the product of $^{16}$O and $^{20}$Ne g.s. ANCs and the theoretical limits for the $^{20}$Ne ANC, are significantly above the most recent transfer reaction ANC value obtained by Shen et al. \cite{Shen2023}, but are consistent with the upper limits of the transfer measurements by Adhikari et al. \cite{Adhikari2017} and Mondal et al. \cite{Mondal2021} within their larger reported uncertainties.

The impact of the $^{16}$O ground state ANC from this work for the $^{12}$C($\alpha$,$\gamma$) reaction can be evaluated by adjusting parameters of the R-matrix calculations presented in deBoer et al. \cite{deBoer2017}. In these fits, we have used the $^{16}$O ground state $\alpha$-ANCs that correspond to the conservative upper and lower limits from Fig.~\ref{fig:16O-ANCs}. Additionally, four other parameters were adjusted: (1) the ANC of the subthreshold 2$^+$ state at 6.92 MeV was increased to the upper limit of the value reported by Avila et al. \cite{Avila2014}; (2) the ANC of the subthreshold 1$^-$ state at 7.12 MeV was adjusted to the nominal value from Avila for consistency; and (3, 4) the widths, $\Gamma_\gamma$ and $\Gamma_\alpha$, of the 1$^-$ state at 9.58 MeV were fixed to their nominal values from literature \cite{Tilley1993}. These changes lead to an excellent fit of the experimental $^{12}$C($\alpha$,$\gamma$)$^{16}$O reaction excitation function \cite{Schurmann2005} below 3.5 MeV in c.m. (see Fig. \ref{fig:sfactor} in supplemental materials).


A decrease between 5\% and 10\% in the total S-factor in the Gamow region at 300 keV is observed when using lower and upper values of the $^{16}$O ground state ANC from this study. Although these changes are non‑negligible, they remain comfortably within the 20\% uncertainty limit at 300 keV quoted by R. deBoer \cite{deBoer2017}.

{\it Conclusion}---The $^{12}\text{C}(^{20}\text{Ne},^{16}\text{O})^{16}\text{O}$ $\alpha$-transfer reaction was performed at the sub-Coulomb energy of 1 MeV/nucleon to measure the square of the product of the $\alpha$-ANCs in the ground states of $^{20}$Ne and $^{16}$O nuclei. Surprisingly, the result of these measurements indicates high degree of $\alpha$-clustering, approaching the maximum, and highlights the key role of the $\alpha$-cluster correlations played in the structure of light nuclei. The main aim of this work was to independently benchmark predictions of theoretical approaches, and to use the combined input of experiment and theory to extract a new model-independent constraint of the $^{16}$O(g.s.) $\alpha$ ANC. All three state-of-the-art microscopic approaches predict well-developed cluster structure for both nuclei, with nearly perfect agreement for $^{20}$Ne(g.s.), while some variation is observed with respect to the $^{16}$O(g.s.) ANC. Using the geometrical mean of the ANCs observable, the theoretical predictions are in good agreement among themselves and with the experimental result, which can now be used as a benchmark for any new theoretical developments. We also show that using the lower and upper limit of the $^{16}$O(g.s.) $\alpha$-ANC, determined in this work, leads to non-negligible changes in the S-factor in the Gamow window for the $^{12}\text{C}(\alpha,\gamma)^{16}\text{O}$ reaction, which, however, remains within the accepted experimental limits due to a compensating effect of other R-matrix fit parameters. Nonetheless, with the new constraint for the $^{16}$O(g.s.) $\alpha$-ANC provided here, it is next important to reduce uncertainties of the 6.92 MeV 2$^+$ $\alpha$-ANC, the width of the 9.58 MeV 1$^-$ resonance, and other resonance parameters to fully constrain the impact at low energies in future R-matrix evaluations.
\\

\begin{acknowledgments}
The authors are very grateful to Dr. S. Yennello, the Director of the Cyclotron Institute at Texas A\&M University, for her interest and support of this work and for agreeing to serve as an independent observer for this double-blind study. We are also grateful to Dr. V.Z. Goldberg and Dr. R.E. Tribble for constructive criticism and useful comments. The experimental work at Texas A\&M University was supported by the U.S. Department of Energy (DOE), Office of Science, Office of Nuclear Science under Award No. DE-FG02-93ER40773 and also by the National Nuclear Security Administration through the Center for Excellence in Nuclear Training and University Based Research (CENTAUR) under Grant No. DE-NA0003841 and No. DE-NA0004150. The theoretical studies utilizing the CN-CI model, conducted by the Florida State University group, were supported by the DOE, Office of Science, Office of Nuclear Physics under Award No. DE-SC0009883. The theoretical research on SA-NCSM, conducted by the Louisiana State University group, was supported by the U.S. Department of Energy (DOE), Office of Science, Office of Nuclear Science under Award No. DE-SC0023532, and by the National Nuclear Security Administration through the Center for Excellence in Nuclear Training and University Based Research (CENTAUR) under DE-NA0004150 (CENTAUR). This work benefited from high performance computational resources provided by LSU (www.hpc.lsu.edu), the NERSC National Energy Research Scientific Computing Center (DE-AC02-05CH11231), and the Frontera computing project (National Science Foundation award OAC-1818253). The NLEFT collaboration was supported by DOE grants (DE-SC0013365, DE-SC0023175, DE-SC0024520, DE-SC0024586), National Science Foundation grants (PHY–2044632), the European Research Council (ERC) under the European Union’s Horizon 2020 research and innovation programme (AdG EXOTIC, grant agreement No. 101018170), the MKW NRW under the funding code NW21-024-A, the JST ERATO Grant No. JPMJER2304, and the Chinese Academy of Sciences through the President’s International Fellowship Initiative (PIFI) (Grant No. 025PD0022). This material is also based upon work supported by the U.S. Department of Energy, Office of Science, Office of Nuclear Physics, under the FRIB Theory Alliance
award DE-SC0013617. Computational resources for the NLEFT collaboration were provided by the Gauss Centre for Supercomputing e.V. (www.gauss-centre.eu) for computing time on the GCS Supercomputer JUWELS at J{\"u}lich Supercomputing Centre (JSC) and special GPU time allocated on JURECA-DC. Resources from the Oak Ridge Leadership Computing Facility were also used, through the INCITE award “Ab-initio nuclear structure and nuclear reactions.” The NLEFT collaboration extends special thanks to Serdar Elhatisari for discussions on NLEFT calculations and to Bing-Nan Lu for developing the lattice interactions used in the calculations presented here. 
\end{acknowledgments}

\bibliography{ANCRefs}

\clearpage

\onecolumngrid
\section{Supplemental Materials}

\subsection{Lattice Effective Field Theory}
For the lattice calculations presented here, we use the ``simple'' lattice interactions introduced in Ref.~\cite{Lu:2018bat} and also used in Ref.~\cite{Lu:2019nbg,Summerfield:2021oex}, with lattice spacing $a = 1.32$~fm. The interaction parameters were tuned to make the $\alpha$ separation energies for $^{16}$O, and $^{20}$Ne as close to empirical values as possible. Reviews of nuclear lattice effective field theory can be found in Ref.~\cite{Lee:2008fa,Lahde:2019npb}. Except for the Coulomb interactions, they are independent of spin and isospin. Nevertheless, the simple interactions still describe the bulk properties of light and medium nuclei with no more than a few percent relative error for the binding energies and radii.  In order to determine the ANCs for $^{16}$O and $^{20}$Ne, we use the pinhole algorithm introduced in Ref.~\cite{Elhatisari:2017eno}.

\begin{figure}[tbh]
    \subfloat[\label{fig:lattice16O:a}]{\includegraphics[width=0.49\columnwidth]{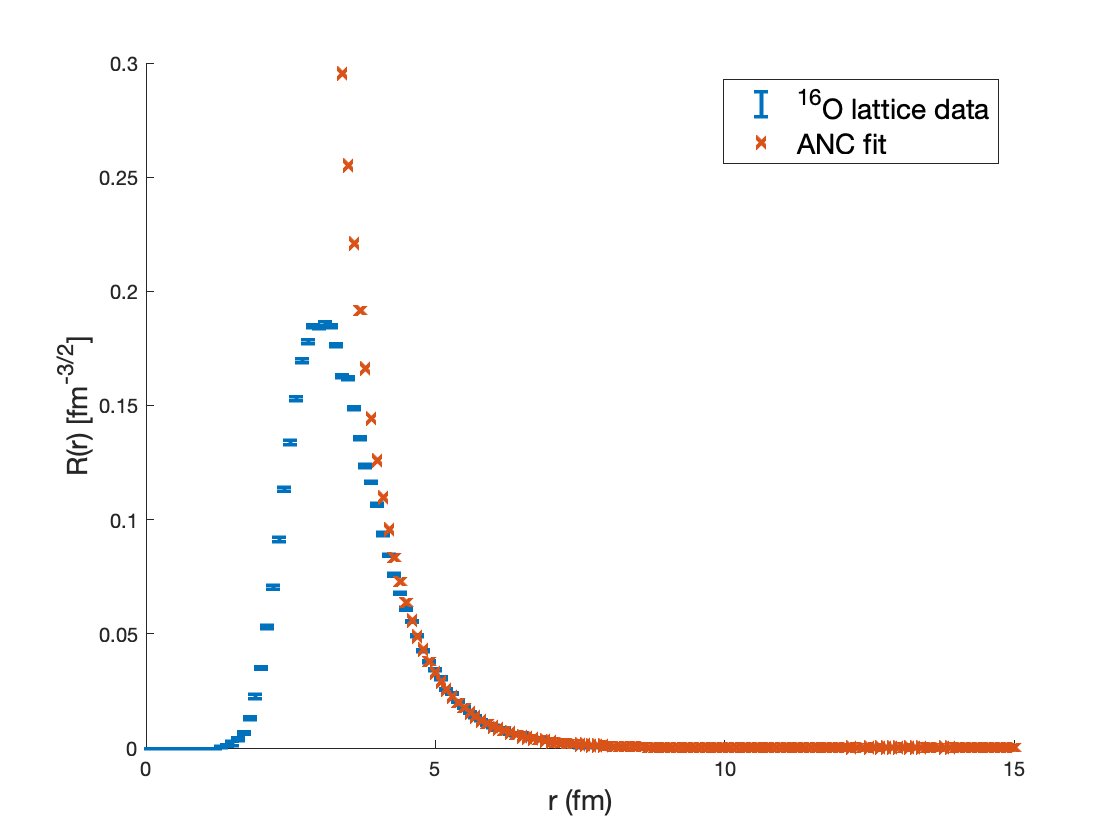}}
    \subfloat[\label{fig:lattice20Ne:b}]{\includegraphics[width=0.49\columnwidth]{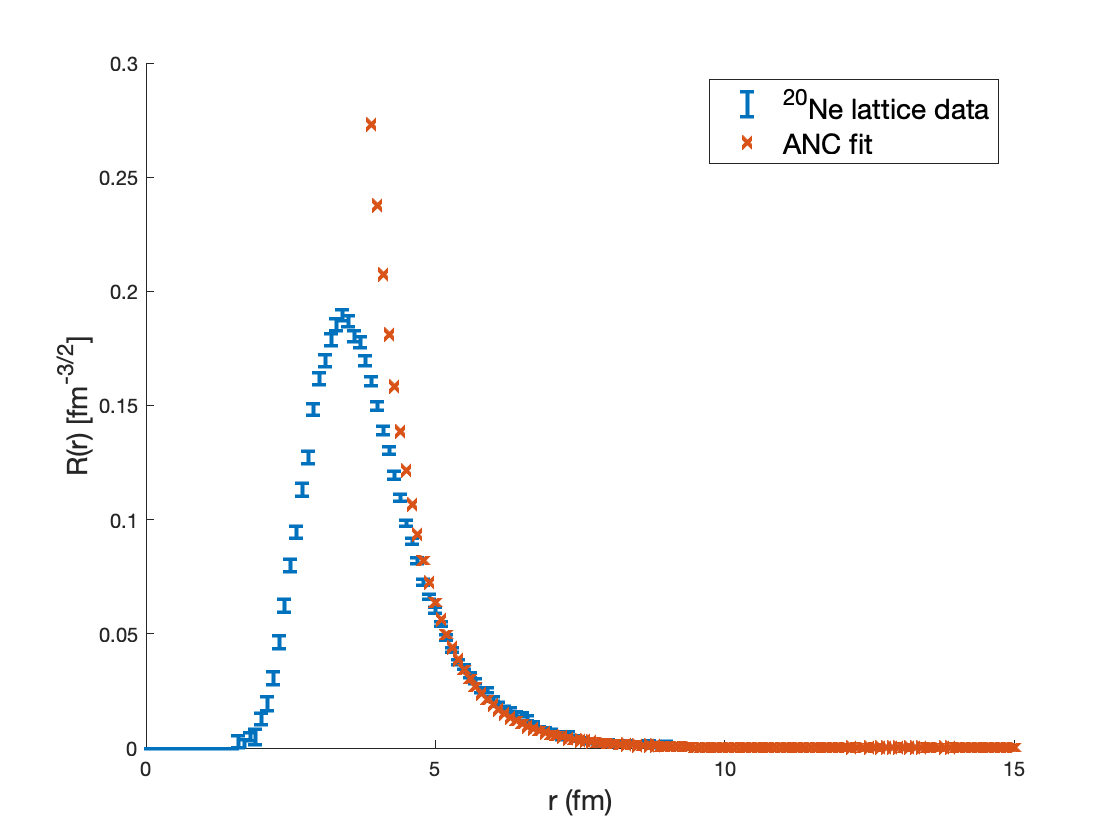}}
    \caption{Lattice results for the radial wavefunction $R(r)$ versus separation distance $r$ between (a) $^{12}$C and $^{4}$He clusters in the $^{16}$O ground state wavefunction, and (b) $^{16}$O and $^{4}$He clusters in the $^{20}$Ne ground state wavefunction. For comparison, we show the best-fit Whittaker functions with ANC values as given in the text.}
    \label{fig:lattice}
\end{figure}

In order to determine the ANC for the ground state of $^{16}$O, we use the pinhole algorithm to produce a classical distribution of the $16$ nucleons with full $A$-body correlations.  We then determine the location of the $^{4}$He cluster furthest away from the center of mass and determine the probability distribution for the relative separation between that $^{4}$He cluster and the remaining $^{12}$C nucleus. From this we extract the tail of the radial wavefunction, and the results are shown in Fig.~\ref{fig:lattice}a. We also show the Whittaker function with the ANC value that produces the best fit to the lattice data. Including stochastic error bars and the uncertainty in the fit, we estimate $380 \pm 80$~fm$^{-1/2}$. The error bar does not include uncertainties due to the nuclear interactions or the extrapolation to infinite Euclidean time and lattice volume.

The ANC calculation for the ground state of $^{20}$Ne is done in an analogous manner. The results are shown in Fig.~\ref{fig:lattice}b and the ANC values are $3800 \pm 950$~fm$^{-1/2}$.

\subsection{Symmetry-adapted No-core Shell Model}
The symmetry-adapted no-core shell model with continuum is an \textit{ab initio} many-body approach that uses chiral effective-field-theory interactions and is reviewed in Refs.~\cite{LauneyDD16,LauneyMD_ARNPS21}. It employs a symmetry-adapted many-body basis to achieve ultra-large model spaces that are imperative for the description of clustering, collectivity, and continuum degrees of freedom \cite{DytrychLDRWRBB20,DreyfussLESBDD20,PhysRevLett.128.202503}. We use a harmonic oscillator (HO) single-particle basis with frequency $\hbar \omega$ and a model space with a cutoff given by the total HO excitations, related to the accessible number of HO shells \cite{DytrychSBDV_PRL07,DytrychLDRWRBB20}. For fixed $\hbar \omega$, HO shells, and inter-nucleon interaction, the results coincide with those of the traditional no-core shell model \cite{NavratilVB00,BarrettNV13}, and become independent of $\hbar \omega$ for an infinite model space, providing a parameter-free prediction.  The matching to the exact Coulomb eigenfunctions at long distances is discussed in Ref.~\cite{DreyfussLESBDD20}. In this study, we calculate $(C^{^{20}{\rm Ne}}_{\alpha,^{16}{\rm O}})^2(C^{^{16}{\rm O}}_{\alpha,^{12}{\rm C}})^2$ for $\hbar \omega=11,\, 13,\, 15$ MeV, and 8-12 (9-13) HO shells for $^{16}$O ($^{20}$Ne), using the NNLO$_{\rm opt}$ chiral potential \cite{Ekstrom13}, the same potential used in Refs.~\cite{DreyfussLESBDD20,PhysRevLett.128.202503}. We find that this ANC product decreases with increasing model spaces for $\hbar \omega=11$ MeV and increases for $\hbar \omega=13$ MeV. Hence, the corresponding Shanks extrapolations to the infinite model space \cite{Shanks55,DytrychLDRWRBB20} provide the upper and lower bounds for the SA-NCSM estimate. These bounds are reported as error bars for the SA-NCSM prediction, hence our errors do not include uncertainties arising from the underlying inter-nucleon interaction.

\begin{figure}[th]
    \centering
    \includegraphics[width=9cm]{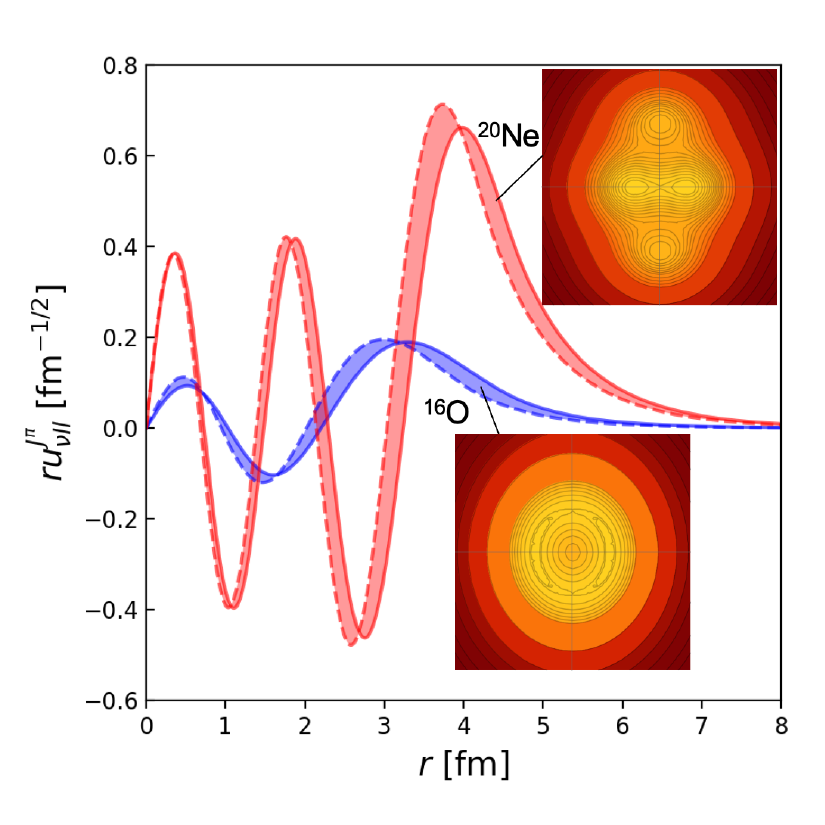}
    \caption{SA-NCSM cluster wavefunction versus the cluster separation $r$ in the $^{16}$O (blue) and $^{20}$Ne (red) ground states for $\hbar \omega=13$  (solid) and 15~MeV (dashed), along with their \emph{one-body} intrinsic densities ($z$-axis versus $r_{xy}$). Lower density in the center of $^{16}$O is consistent with a tetrahedron structure; the more loosely structure of $^{20}$Ne exhibits a more pronounced surface $\alpha$ clustering.}
    \label{fig:sancsm}
\end{figure}

\subsection{Cluster Nucleon Configuration Interaction Model}
The Cluster Nucleon Configuration Interaction technique used in a series of theoretical studies stems from several previous works \cite{kravvaris2017Constructing, kravvaris2017Study, kravvaris2019Clustering, volya2015Nuclear}. 
The technique is built on a harmonic oscillator (HO) basis that is used to describe each individual nuclear system, starting from nucleonic degrees of freedom, as in the traditional no-core shell model \cite{navratil2000Largebasis}. The treatment is extended to include clusters whose motion is also described using an expansion in the harmonic oscillator wave functions. This approach maintains full fermionic antisymmetry and translational and rotational invariance. The entire model represents a configuration interaction treatment involving traditional Slater determinant configurations and configurations describing cluster dynamics. The matching of cluster dynamics expanded in the HO basis with the remote asymptotic behavior and the proper normalization required for ANC are discussed in Refs. \cite{kravvaris2017Study, kravvaris2019Clustering}. We use three different Hamiltonians within this approach: the phenomenological PSDU Hamiltonian \cite{utsuno2011Multiparticlemultihole}, the newer phenomenological no-core effective FSU Hamiltonian \cite{lubna2020Evolution}, and the no-core JISP Hamiltonian from Ref. \cite{shirokov2007Realistic}. We assess the theoretical uncertainties from the differences.

\subsection{Experiment}

\begin{figure}[tbh]
    \subfloat[\label{fig:20MeV_tofx1:a}]{\includegraphics[width=0.49\columnwidth]{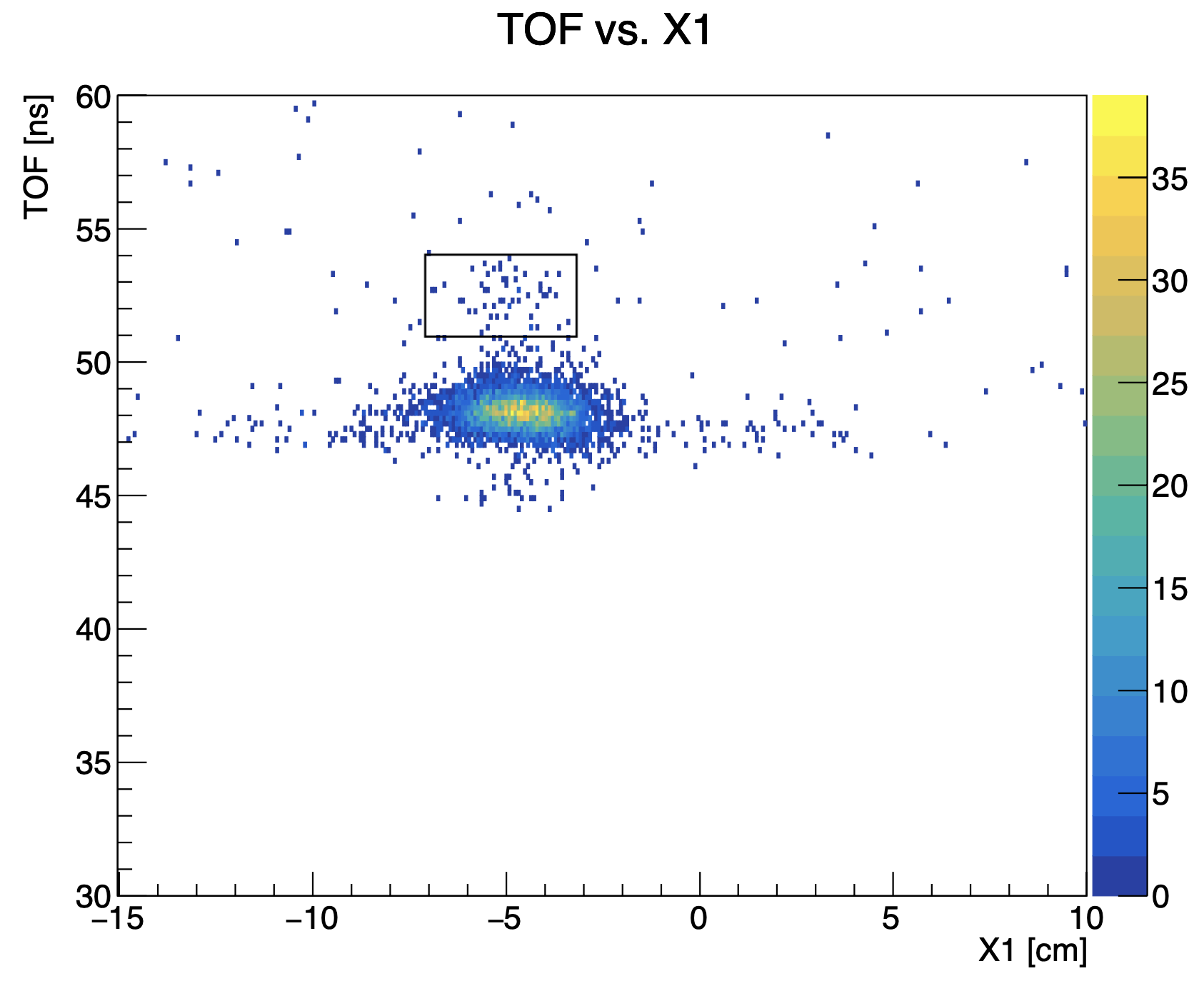}}
    \subfloat[\label{fig:22MeV_tofx1:b}]{\includegraphics[width=0.49\columnwidth]{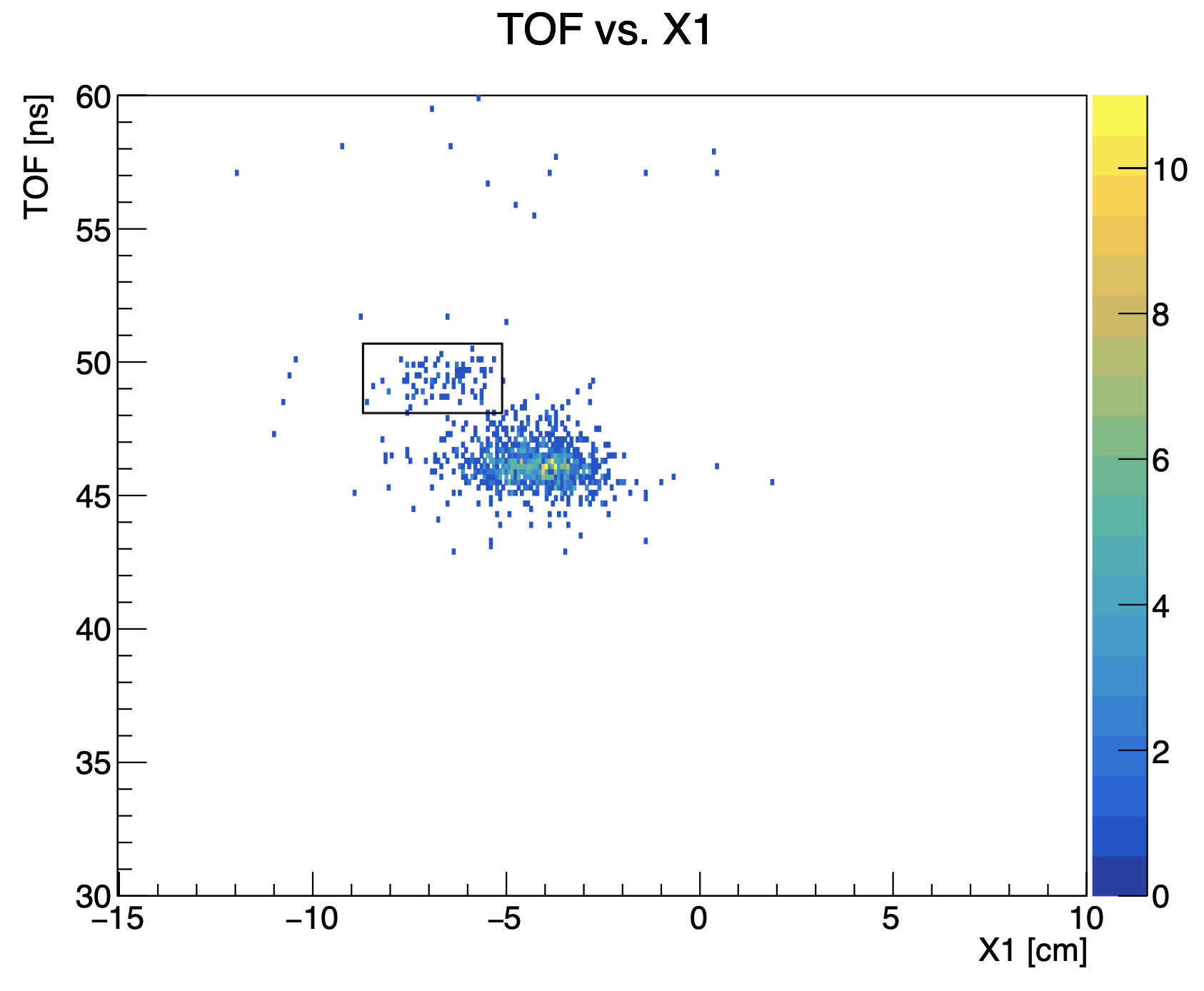}}
    \caption{TOF between the two PPACs versus the X position of the first PPAC for an MDM angle of 5$^\circ$ and beam energies of (a) 20~MeV and (b) 22~MeV. The black boxes indicate $^{16}$O events from the $\alpha$-transfer reaction.}
    \label{fig:ppac-data}
\end{figure}

The $^{12}$C($^{20}$Ne,$^{16}$O)$^{16}$O $\alpha$-transfer reaction was measured at two different beam energies of 20 and 22 MeV. Data was taken at 5$^{\circ}$ for both beam energies, with an additional measurement made at an angle of 15$^{\circ}$ for the 22 MeV beam. Figure \ref{fig:ppac-data} shows the 2D particle identification plots of the $^{16}$O detected in TexPPAC. The dominant peak in both spectra corresponds to $^{13}$C$^{5^+}$ from the natural carbon target, which has a similar magnetic rigidity to the $^{16}$O$^{6^+}$ from the transfer reaction. To determine the total number of background subtracted $^{16}$O counts, we used a multi-Gaussian fit. For normalization of the $\alpha$-transfer cross section, $^{12}$C+$^{20}$Ne elastic scattering was measured in TexPPAC at both beam energies and angles.

\begin{figure}
    \centering
    \includegraphics[width=0.7\linewidth]{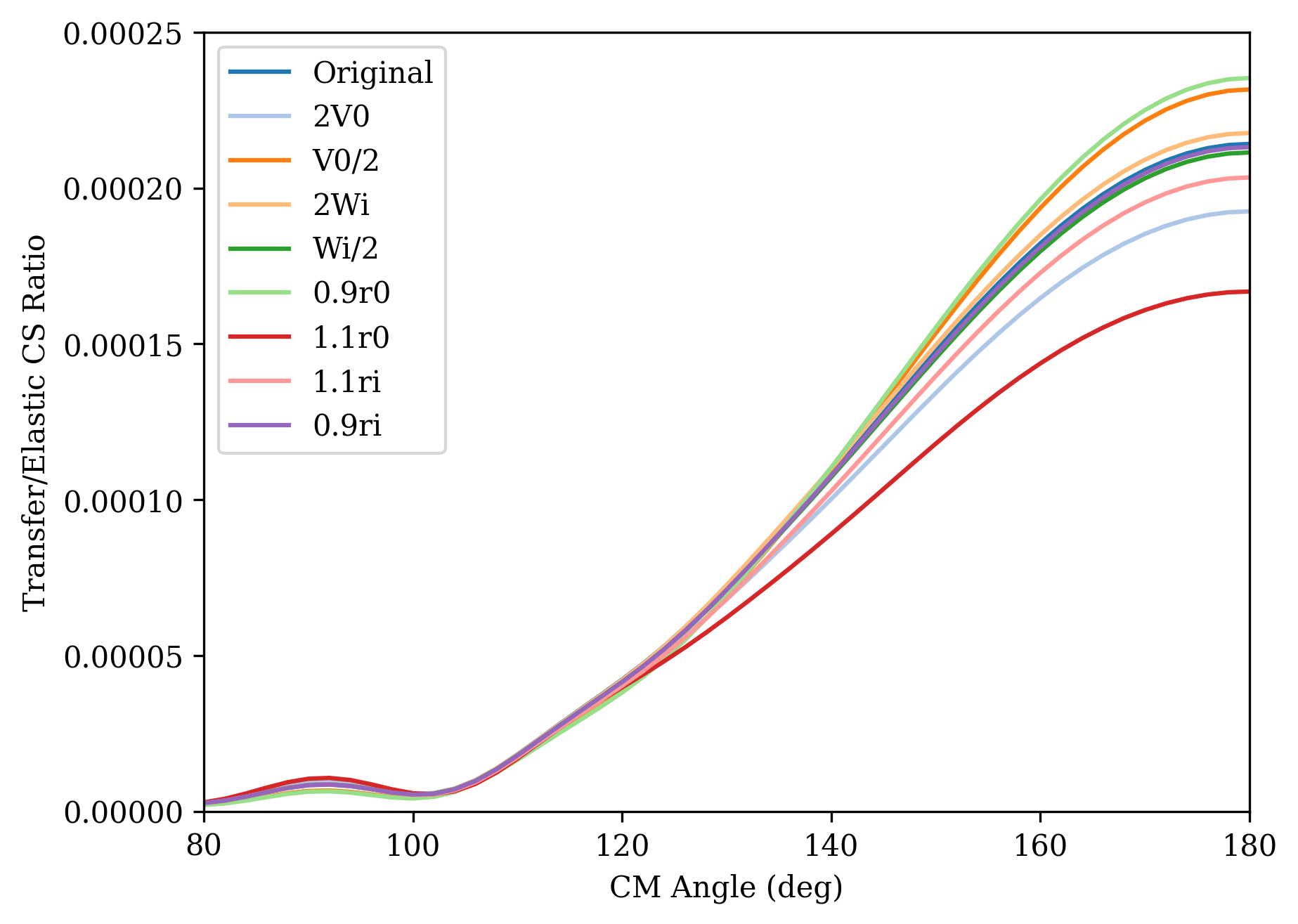}
    \caption{The 20 MeV transfer to elastic cross section ratio plotted against c.m. angle for variations in the $^{16}$O+$^{16}$O potential from \cite{Wu1984}. V0 (Wi) and r0 (ri) represent the real (imaginary) potential depth and radius.}
    \label{fig:cs_ratio}
\end{figure}

The experimental ANC product is determined by multiplying the single-particle ANCs ($b$) with the experimental and DWBA $\alpha$-transfer differential cross sections normalized to $^{12}$C+$^{20}$Ne elastic scattering:
\begin{equation}
(C_{\alpha,^{16}O}^{^{20}Ne})^2(C_{\alpha,^{12}C}^{^{16}O})^2 = (b_{\alpha,^{16}O}^{^{20}Ne})^2(b_{\alpha,^{12}C}^{^{16}O})^2 \left(\frac{\left(\frac{d\sigma}{d\Omega}\right)^{\text{transfer}}_{\text{exp}}}{\left(\frac{d\sigma}{d\Omega}\right)^{\text{elastic}}_{\text{exp}}}\right)
    \left(\frac{\left(\frac{d\sigma}{d\Omega}\right)^{\text{elastic}}_{\text{DWBA}}}{\left(\frac{d\sigma}{d\Omega}\right)^{\text{transfer}}_{\text{DWBA}}}\right),
     \label{eq_ANC_Product}
\end{equation}
where single-particle ANCs are obtained with FRESCO \cite{Thompson1988}. Utilizing Eq. \ref{eq_ANC_Product}, we achieve consistent results between all measurements conducted at 20 and 22 MeV. The dependence of the 20 MeV result on the $^{16}$O+$^{16}$O OMP parameters of the DWBA calculation is shown in Fig. \ref{fig:cs_ratio}, where we conclude a total theoretical uncertainty of 20\%. Due to the greater sensitivity to these parameter variations at 22~MeV, only the results from the lower-energy (20 MeV) measurement are considered in the final analysis.

\subsection{$^{12}\text{C}(\alpha,\gamma)^{16}\text{O}$ S-factor}

\begin{figure}
    \centering
    \includegraphics[width=0.7\linewidth]{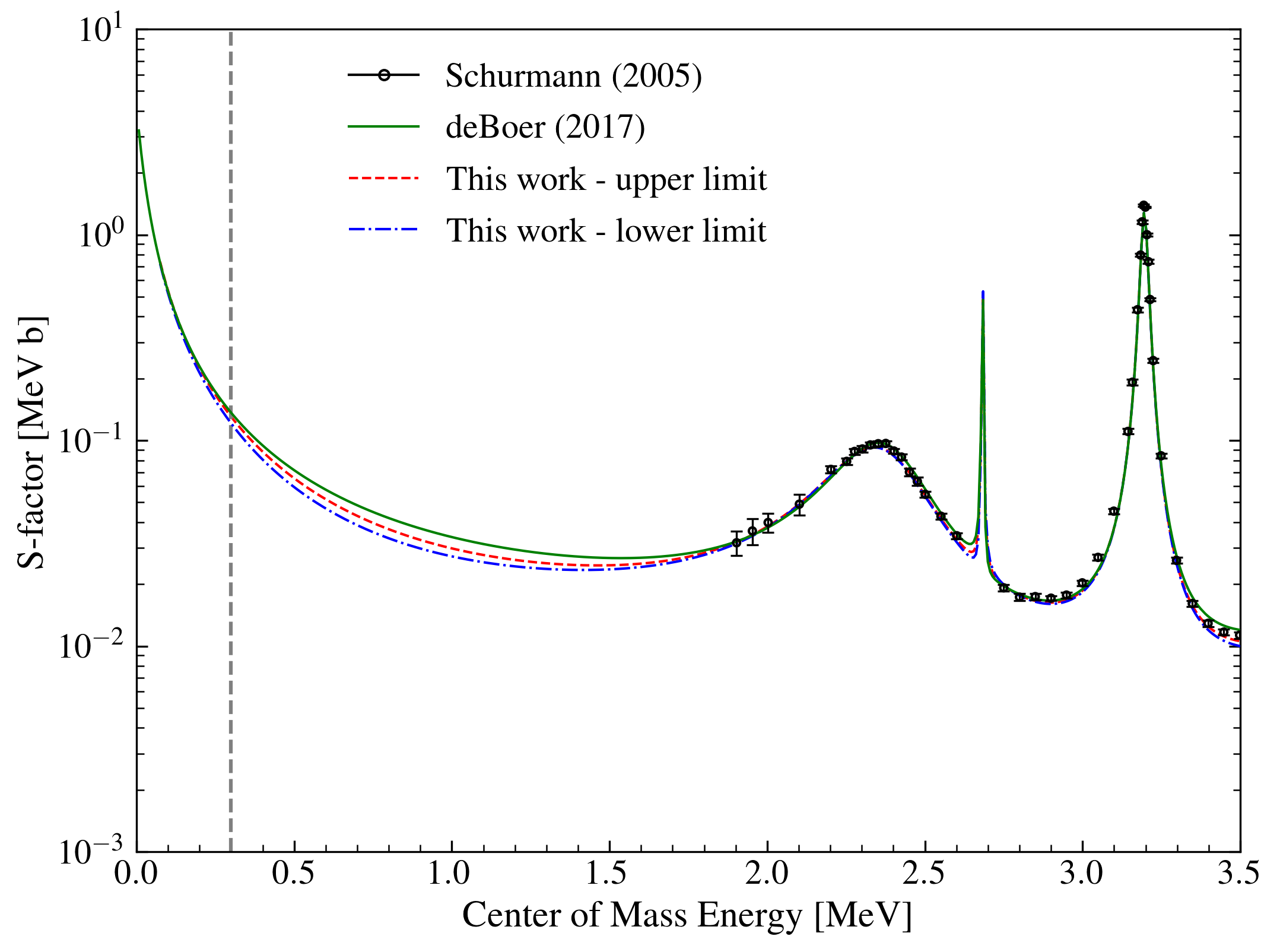}
    \caption{Comparison of R-matrix calculations of the total S-factor of the $^{12}$C($\alpha,\gamma$)$^{16}$O reaction. The solid green line shows the R-matrix fit of deBoer et al. \cite{deBoer2017}. The dashed red and dashed-dotted blue lines denote the upper and lower limits of the total S-factor as determined by this work, respectively. The black points are the experimental data of the total S-factor given by Sh\"urmann et al. \cite{Schurmann2005}. The grey dashed line indicates the Gamow energy at 300 keV.}
    \label{fig:sfactor}
\end{figure}

The R-matrix code AZURE2 \cite{azuma2010} was used to establish the upper and lower bounds for the $^{12}$C($\alpha,\gamma$)$^{16}$O S-factor in the Gamow energy region around 300 keV based on the high and low values of the $^{16}$O(g.s.) ANC. Fig. \ref{fig:sfactor} shows a comparison of the S-factor fits from this work with the fit of deBoer \cite{deBoer2017} (represented by the solid green curve). The red dashed and blue dot-dashed curves represent the upper and lower limit fits from this work, respectively. The parameters which were adjusted to reproduce the fit are described in the main text. The most noticeable differences in the fits occur in the off-resonance regions between $0.5~\text{MeV} < E_{CM} < 2~\text{MeV}$, while the experimental data below 3.5 MeV remain well reproduced. It is important to emphasize that the four parameters we adjusted remain within their respective experimental uncertainty limits.

\end{document}